\documentclass[nofootinbib,floatfix,superscriptaddress]{revtex4}

\usepackage{graphicx}
\usepackage{epsfig}
\usepackage{bm}
\usepackage[T1]{fontenc}
\usepackage[latin9]{inputenc}
\usepackage{amssymb}
\usepackage{float}
\usepackage{amsmath}
\usepackage{dcolumn}
\usepackage{cancel}
\usepackage[colorlinks]{hyperref}
\usepackage[usenames,dvipsnames]{color}
\hypersetup{
     breaklinks=true,
    pdfstartview={FitH},    
    colorlinks=true,       
    linkcolor=blue,          
    citecolor=red,        
    filecolor=magenta,      
    urlcolor=blue,           
    anchorcolor=green,      
    linktocpage=true
}

\newcommand{\be}{\begin{equation}}
\newcommand{\ee}{\end{equation}}

\begin{document}

\title{Lagrangian reconstruction of Barrow holographic dark energy in interacting tachyon model}

\author{G.~G.~Luciano}
\email{giuseppegaetano.luciano@udl.cat}
\affiliation{Applied Physics Section of Environmental Science Department,  Escola Polit\`ecnica Superior, Universitat de Lleida, Av. Jaume
II, 69, 25001 Lleida, Spain}

\author{Y. Liu}
\email{yang.liu@nottingham.ac.uk}
\affiliation{School of Physics and Astronomy, University of Nottingham, Nottingham NG7 2RD, UK}

\date{\today}

\begin{abstract}
We consider a correspondence between the tachyon dark energy
model and Barrow holographic dark energy (BHDE). The latter is a modified scenario based on the application of the holographic principle with Barrow entropy instead of the usual Bekenstein-Hawking one. We reconstruct the dynamics of the tachyon scalar field $T$ in a curved Friedmann-Robertson-Walker Universe both in the presence and absence of interactions between dark energy and matter. As a result, we show that the tachyon field exhibits a non-trivial dynamics. In a flat Universe, $\dot T^2$ must always be vanishing, independently of the existence of interaction. This implies $\omega_D=-1$ for the equation-of-state parameter, which in turn can be used for modeling the cosmological constant behavior. On the other hand,  for a non-flat Universe and various values of Barrow parameter, we find that $\dot T^2$ decreases monotonically for increasing $\cos(R_h/a)$ and $\cosh(R_h/a)$, where $R_h$ and $a$ are the future event horizon and the scale factor, respectively. Specifically, $\dot T^2\ge0$ for a closed Universe, while $\dot T^2<0$ for an open one, which is physically not allowed. We finally comment on the inflation mechanism and Trans-Planckian Censorship Conjecture in BHDE and discuss observational consistency of our model.
\end{abstract}

 \maketitle

\section{Introduction}
\label{Intro}

Experimental evidences from Supernova SNIa,
Baryon acoustic oscillations and gravitational waves
have definitely proved that our Universe is expanding
at an accelerated rate. In spite of enormous effort, a fully consistent explanation for the origin of this behavior is missing. Among the various mechanisms, the existence of an unknown form of energy (Dark Energy, DE) affecting the Universe on large scales is the 
most widely accepted proposal. But yet, the nature of DE remains quite elusive. The possibility that DE is modeled by the cosmological constant 
acting as source of vacuum energy has been originally 
considered as natural way out of the DE puzzle~\cite{Sahni:1999gb,Peebles:2002gy}. However, this scenario is at odds 
with our field theoretical understanding of the quantum properties of vacuum, 
thus requiring further investigation. Along this line, a plethora of
DE models have been put forward over the years~\cite{Ratra:1987rm,Frieman:1995pm,Turner:1997npq,Caldwell:1997ii,Armendariz-Picon:2000nqq,Armendariz-Picon:2000ulo,Caldwell:1999ew,Caldwell:2003vq,Nojiri:2003vn,Feng:2004ad,Guo:2004fq,Elizalde:2004mq,Nojiri:2005sx,Deffayet:2001pu,DagoTsallis,Capoz1,Capol1,Capol2,Lamb1,Lamb2}.

An interesting model to account for the nature of DE
is the so called Holographic Dark Energy (HDE)~\cite{Cohen:1998zx,Horava:2000tb,Thomas:2002pq,Li:2004rb,Bamba:2012cp,Ghaffari:2014pxa,Wang:2016och}, which emerges within quantum gravity framework.
The main ingredient of this approach is the 
holographic principle, according to which the description of a volume of space can be thought of as encoded on a lower-dimensional boundary surface to the region.  
In~\cite{tHooft:1993dmi,Susskind:1994vu} it has been pointed
out that effective local quantum field theories over-count the number of 
independent degrees of freedom, predicting that entropy scales
extensively ($S\sim L^3$) for systems of size $L$ with UV cutoff $\Lambda$. 
Later on, a solution to this problem has been provided in~\cite{Cohen:1998zx}, 
where it has been argued that the total energy of a system with size $L$
should not exceed that of an equally sized black hole, i.e. $L^3\rho_\Lambda\le L M_p^2$. Here, $M_p=(8\pi G)^{-1/2}$ is the reduced Planck mass, while $\rho_\Lambda$ denotes
the quantum zero-point energy density caused by the UV cutoff $\Lambda$
(we are working in natural units $\hslash=c=1$). The inequality is saturated
for the largest value of $L$. In this context, 
the holographic dark energy density is obtained as 
\be
\label{HDE}
\rho_\Lambda=\frac{3c^2 M_p^2}{L^2}\,,
\ee
where $c$ is a dimensionless constant and the
factor $3$ has been introduced for convenience. 

Cosmological applications of the holographic principle and HDE 
have been largely considered in literature. As an example, it was analyzed by~\cite{Enqvist:2004ny} that the consequence of excluding
those degrees of freedom of the system that will never be observed by the effective field theory results into an IR cutoff $L$ at the future event horizon.
In a DE dominated Universe, such an horizon is then predicted to tend toward a constant value of the order $H_0^{-1}$,
with $H_0$ being the present {Hubble rate}~\cite{Setare:2007hq}. 
Furthermore, the issue of assuming the apparent (Hubble) horizon
$R_A=1/H$ as IR cutoff in a flat Universe has been examined in~\cite{Hsu:2004ri, Sriva}.

Despite the intensive study, the shortcomings of the HDE 
in describing the history of a flat Friedmann-Robertson-Walker (FRW) Universe have prompted tentative changes to this approach. For instance, HDE has been used to address the DE problem in Brans-Dicke Cosmology~\cite{Gong:1999ge,Kim:2005gk,Nojiri:2005pu,Setare:2006yj,Banerjee:2000gt,Xu:2008sn,Khodam-Mohammadi:2014wla}
by considering different IR cutoffs~\cite{Wang:2016och,Ghaffari:2014pxa,Xu:2008sn} or and/or generalized entropies~\cite{Tavayef:2018xwx,Saridakis:2018unr,Nojiri:2019skr,GineLuc,Saridakis:2020zol,Moradpour:2020dfm,Drepanou:2021jiv,Hernand,Nojiri:2021jxf,LucKan,GhaffariLuc,Nojiri:2022aof,Nojiri:2022dkr,Luciano:2022ffn,Luciano:2022hhy,LucianoPLB}. 
In particular, the latter path has led to promising models, 
such as Tsallis~\cite{Tavayef:2018xwx,Saridakis:2018unr,Nojiri:2019skr,OdCor,GineLuc},
Barrow~\cite{Saridakis:2020zol,Nojiri:2021jxf,GhaffariLuc} and Kaniadakis~\cite{Drepanou:2021jiv,Hernand,LucKan} holographic dark energy, the latter two being motivated by quantum gravitational and relativistic considerations, respectively.  

While exhibiting a richer phenomenology comparing to standard Cosmology, HDE with generalized entropies is often based
on \emph{ad hoc} deformations of the entropy-area law, 
which might somehow question the relevance of this model in
improving our knowledge of the Universe at very fundamental level. 
In the absence of solid empirical guidelines, 
valuable hints can be gained by looking at Noether symmetries
of the underlying theoretical framework, which are 
notoriously linked to physical conserved quantities (see also~\cite{Bajardi:2021tul,Acunzo:2021gqc} for other recent usage of Noether symmetries in cosmology and gravitation).
In order to export Noether's theorem to the present analysis, 
a reformulation of extended HDE is needed
using the Lagrangian language. Toward this end, 
a pathway is to consider
reconstruction scenarios, i.e. to compare the 
relative energy density of extended HDE
and other solid cosmological models to find the reconstructed 
action which reproduces the whole cosmic history of the Universe. Reconstruction paradigm could be particularly inspiring
for Barrow Holographic Dark Energy as a preliminary 
attempt to formulate the effective action of cosmological model in a quantum gravity-oriented picture.  

Along this line and motivated by the analysis of~\cite{Setare:2007hq}, in~\cite{Liu:2021heo}
it has been shown that Tsallis holographic description of DE (THDE) is non-trivially intertwined with tachyon dark energy model~\cite{Liu:2021heo}. {In this regard, we would like to remark that 
the tachyon field has been proposed as a possible candidate for dark energy. In particular, a rolling tachyon has a peculiar equation of state parameter that interpolates between the values $-1$ and $0$~\cite{Gibbons2002}. Therefore, the tachyon can be realized as 
a suitable candidate for the high energy inflation~\cite{Mazum} and at the same time as a source of dark
energy depending on the form of the tachyon potential~\cite{Padmanab}. Not least, such a model is mathematically less cumbersome than scalar-tensor theory, thus allowing for a more 
direct and intuitive interpretation of results.}
In~\cite{Liu:2021heo}
a correspondence between the tachyon field and THDE  has been 
established based on the reconstruction of
the dynamics of the tachyon field in THDE. In recent years, 
more applications have been analyzed 
in $f(R)$~\cite{EPL}, $f(R,T)$~\cite{Chinese}, 
$f(G,T)$~\cite{fgtgrav}, teleparallel~\cite{Wahe}, Brans-Dicke~\cite{GhaffariBD}, logarithmic Brans-Dicke~\cite{LogBD} and 
Saez-Ballester~\cite{Santhi,Santhibis} theories, among others.

Starting from the above premises, in this work we explore more in-depth the connection between the tachyon dark energy model and HDE. We frame our analysis in the context of HDE based on Barrow entropy~\cite{Saridakis:2020zol}. {The ensuing scenario is typically named Barrow Holographic Dark Energy (BHDE) and arises from the application of the holographic principle at a cosmological framework, but employing Barrow entropy~\cite{Barrow:2020tzx} instead of the standard Bekenstein-Hawking one.}
We analyze the case of a non-flat FRW Universe for interacting dark energy. 
Since scalar fields are generally conjectured to have driven inflation
in the very early Universe, we then study the inflation mechanism in our BHDE model. We find an analytical solution for 
the slow-roll parameters, the scalar
spectral index and the tensor-to-scalar ratio. We also compare our findings with recent results in the literature. 

The remainder of the work is structured as follows:
in the next Section we introduce BHDE. Section~\ref{Tachnonflat} 
is devoted to analyze the correspondence between the tachyon dark energy and BHDE in a non-flat FRW Universe. In Sec.~\ref{infl} we discuss inflation in BHDE, while conclusions and outlook are summarized in 
Sec.~\ref{Conc}. 

\section{Barrow holographic dark energy}
\label{BHDE}
Let us briefly review the basics of BHDE. 
We consider the four-dimensional Friedmann-Robertson-Walker (FRW) metric
\be
\label{FRW}
ds^2\,=\,-dt^2+a^2(t)\left(\frac{dr^2}{1-k\,r^2}+r^2d\Omega^2\right),
\ee
of scale factor $a(t)$ and spatial curvature $k=0,1,-1$ for a flat, closed and open Universe, respectively. 

We use the definition~\eqref{HDE} for the holographic dark energy in standard Cosmology and 
assume~\cite{Setare:2007hq}
\be
\label{ldef}
L(t)\,=\,a(t) r(t)\,,
\ee 
where $r(t)$ is the (time-dependent) radius  that is relevant to the future event horizon of the Universe. Since
\begin{eqnarray}
\label{r}
\int_0^{r_1}\frac{dr}{\sqrt{1-k r^2}}&=&\frac{1}{\sqrt{|k|}}\sin\mathrm{n}^{-1}(\sqrt{|k|}r_1)\\[2mm]
&=&\left\{\begin{array}{rcl}
\nonumber
&&\hspace{-0.4cm}\sin^{-1}(\sqrt{k}\,r_1)/\sqrt{k}, \hspace{1.07cm}k=1,  \\[3mm]
\vspace{2mm}
\nonumber
&&\hspace{-0.4cm}r_1,\hspace{3.35cm} k=0,  \\[1mm]
\nonumber
&&\hspace{-0.4cm}\sinh^{-1}(\sqrt{|k|}r_1)/\sqrt{|k|},\,\,\,\,\,\,\,k=-1,
\end{array}\right.
\end{eqnarray}
we easily obtain
\be
\label{lbis}
L(t)\,=\,\frac{a(t)\sin\mathrm{n}\left[\sqrt{|k|}R_h(t)/a(t)\right]}{\sqrt{|k|}}\,,
\ee
where $R_h$ is the future event horizon
given by~\cite{Setare:2007hq} 
\be
R_h\,=\,a\int_t^{\infty}\frac{dt}{a}\,=\,a\int_a^{\infty}\frac{da}{Ha^2}\,,\quad\,\, H=\frac{\dot a}{a}\,.
\ee

HDE relies on the holographic principle, which 
asserts that the number of degrees of freedom describing the physics of any quantum gravity system \emph{i}) scales as the bounding surface (rather than the volume) of the system and \emph{ii}) should be constrained by an infrared cutoff~\cite{tHooft:1993dmi,Susskind:1994vu}. This is in tune
with Bekenstein-Hawking (BH) relation $S_{BH}=A/A_0$ for black holes, where $S_{BH}$ and $A$ denote the entropy and area of the black hole, respectively, while $A_0=4G$ its Planck area. Recently, deformations of this relation have been proposed to
take account of quantum~\cite{Barrow:2020tzx,Tsallis:1987eu,Tsallis:2012js} and/or relativistic~\cite{Kaniadakis:2002zz} effects. 
In particular, in~\cite{Barrow:2020tzx} it has been argued
that quantum gravity may introduce intricate, fractal features
on the black hole horizon, leading to the modified area law
\be
\label{Barrow}
S_\Delta\,=\,\left(\frac{A}{A_0}\right)^{1+\Delta/2}.
\ee
Deviations from BH entropy are quantified by the
exponent $0\le\Delta\le1$, with $\Delta=0$ giving the BH
limit, while $\Delta=1$ corresponding to the maximal horizon
deformation. We emphasize that although this relation 
resembles Tsallis entropy in non-extensive statistical thermodynamics~\cite{Tsallis:1987eu,Tsallis:2012js}, the origin and motivation 
underlying Eq.~\eqref{Barrow} are completely different. Cosmological implications of Barrow entropy have been recently studied in
the context of Big Bang Nucleosynthesis~\cite{Barrow:2020kug}, Baryogenesis~\cite{Luciano:2022pzg} and tests of gravity theories 
from observations of Sagittarius A*~\cite{Vagnozzi}, among others. 
The possibility of a running $\Delta$ has also been considered in~\cite{Running}.

Strictly speaking, Eq.~\eqref{Barrow} has been
formulated for black holes. However, it 
is known that in any gravity theory one can 
consider the entropy for the Universe 
horizon in the same form as the black hole entropy, the only adjustment being the replacement of the black hole horizon radius
with the apparent horizon radius. This is at the heart
of the various generalizations of HDE with 
modified entropy laws (see, e.g.~\cite{Tavayef:2018xwx,Saridakis:2018unr,Saridakis:2020zol,Moradpour:2020dfm,Drepanou:2021jiv}).

Now, in~\cite{Cohen:1998zx} Cohen et al. have proposed
the following inequality between the entropy, the IR ($L$)
and UV ($\Lambda$) cutoffs for a given system in an effective local quantum field theory 
\be
L^3 \Lambda^3\,\le\, S_{max}\simeq\,S_{BH}^{3/4}\,.
\ee
If we use for the entropy the modified expression~\eqref{Barrow}, we have
\be
\Lambda^4\,\le\,\left(2\sqrt{\pi}\right)^{2+\Delta}\frac{L^{\Delta-2}}{A_0^{1+\Delta/2}}\,,
\ee
where $\Lambda^4$ denotes the vacuum energy density, 
i.e. the energy density of DE ($\rho_D$) in the HDE hypothesis~\cite{Guberina:2006qh}.

By using the above inequality, Barrow holographic dark energy density
can be proposed as
\be
\label{brhod}
\rho_D\,=\,C L^{\Delta-2}\,,
\ee
where $C$ is an unknown parameter with dimensions $[L]^{-2-\Delta}$.
It is worth noticing that for $\Delta=0$, the above relation
reduces to the standard HDE~\eqref{HDE}, provided that $C=3c^2M_p^2$. On the other hand, in the case where
deformations effects switch on ($\Delta\neq0$), BHDE
departs from the standard HDE, leading to different cosmological
scenarios~\cite{Saridakis:2020zol}. 

Following the standard literature, we now define the critical
energy density $\rho_{cr}$ and the curvature energy density
$\rho_k$ as
\be
\rho_{cr}\,=\, 3 M_p^2 H^2\,,\qquad \rho_k\,=\,\frac{3k}{8\pi G a^2}\,.
\ee
We also introduce the three fractional energy densities 
\begin{eqnarray}
\label{Om}
\Omega_m&=&\frac{\rho_m}{\rho_{cr}}\,=\,\frac{\rho_m}{3M_p^2H^2}\,,\\[2mm]
\label{OD}
\Omega_D&=&\frac{\rho_D}{\rho_{cr}}\,=\,\frac{C}{3M_p^2H^2}L^{\Delta-2}\,,\\[2mm]
\label{Ok}
\Omega_k&=&\frac{\rho_k}{\rho_{cr}}\,=\,\frac{k}{H^2a^2}\,,
\end{eqnarray}
{where $\rho_m$ 
is the matter energy density. }
{In particular, by setting $L=H^{-1}$ as in~\cite{Sriva,Tavayef:2018xwx,Hamedan,AlMamon,GhafIR,SheIR,Boul} we obtain}\footnote{There are several choices for the IR cutoff $L$. Following~\cite{Tavayef:2018xwx}, here we resort to the simplest one $L=H^{-1}$.
Other possible choices are the particle horizon, the future event horizon, the GO cutoff~\cite{GO} or combination thereof. However, in these cases one must generally resort to numerical evaluation to study the cosmological evolution of the model~\cite{Saridakis:2018unr}. Since we are interested in extracting analytical solutions and given the degree of arbitrariness in the selection of the most reliable description of dark energy, we leave the analysis of dark energy models with different IR cutoffs for future investigation.}
\be
\Omega_D\,=\,\frac{C}{3M_p^2\,H^{\Delta}}\,.
\ee

From Eqs.~\eqref{ldef} and~\eqref{r}, one can derive the following
expression for the time derivative of $L$~\cite{Liu:2021heo}
\be
\label{dotL}
\dot{L}\,=\,HL+a\dot r\, =\, 1-\frac{1}{\sqrt{|k|}}\cos\mathrm{n}(\sqrt{|k|}R_h/a)\,,
\ee
where we have defined~\cite{Setare:2007hq}
\be
\label{cosn}
\frac{1}{\sqrt{|k|}}\cos \mathrm{n}(\sqrt{|k|}x)=\left\{\begin{array}{rcl}
&&\hspace{-0.4cm}\cos (x)\,, \hspace{0.6cm}k=1,  \\[3mm]
\vspace{2mm}
&&\hspace{-0.4cm}1\,,\hspace{1.37cm} k=0,  \\[1mm]
&&\hspace{-0.4cm}\cosh(x),\,\,\,\,\,\,\,\hspace{0.5mm}k=-1.
\end{array}\right.
\ee

Now, for a flat FRW Universe filled by non-interacting BHDE and pressureless DM, the first Friedmann equation takes the form
\be
\label{ffe}
H^2\,=\,\frac{1}{3M_p^2}\left(\rho_D+\rho_m\right),
\ee
which, by use of Eqs.~\eqref{Om} and~\eqref{OD}, can be
rewritten as
\be
\label{sumOm}
\Omega_m+\Omega_D\,=\,\Omega_D(1+u)\,=\,1\,,
\ee
where 
\be
\label{u}
u=\frac{\rho_m}{\rho_D}=\frac{\Omega_m}{\Omega_D}\,.
\ee 
Since BHDE does not interact with other parts of cosmos (DM), 
the conservation equations of dust and THDE read
\begin{eqnarray}
\label{c1}
&&\dot\rho_m+3H\rho_m\,=\,0\,,\\[2mm]
&&\dot\rho_D+3H\rho_D(1+\omega_D)\,=\,0\,,
\label{c2}
\end{eqnarray}
where we have denoted by $\omega_D=p_D/\rho_D$ and $p_D$
the equation of state parameter and pressure of THDE, respectively.
{From Eq.~\eqref{c1}, we obtain $\rho_m=\rho_{m,0}/a^3$, where $\rho_{m,0}$ is the present matter energy density.}

Deriving Eq.~\eqref{ffe} respect to time and using
the continuity equations~\eqref{c1} and~\eqref{c2}, after
some algebra we are led to
\be
\label{dotH1}
\frac{\dot H}{H^2}\,=\,-\frac{3}{2}\left(1+\omega_D+u\right)\Omega_D\,.
\ee
Likewise, by plugging Eq.~\eqref{brhod} into~\eqref{c2}, 
we find
\be
\label{dotH2}
\frac{\dot H}{H^2}\,=\,(1+\omega_D)\frac{3}{\Delta-2}\,,
\ee
which gives, by comparison with Eq.~\eqref{dotH1}
\be
\label{omegad}
\omega_D\,=\,\frac{u(2-\Delta)\Omega_D}{2-(2-\Delta)\Omega_D}-1\,.
\ee
With the aid of Eq.~\eqref{sumOm}, this finally yields\footnote{Unlike THDE model, where $\omega_D$ is divergent for $\delta_T<1$ and $\Omega_D=1(2-\delta_T)$ ($\delta_T$ is the Tsallis exponent), BHDE is well-defined for any value of $0\le\Delta\le1$. }
\be
\label{omegaD}
\omega_D\,=\,\frac{\Delta}{(2-\Delta)\Omega_D-2}\,.
\ee

On the other hand, if there exists an interaction of the type
\be
\label{intercoup}
Q=3b^2H(\rho_m+\rho_D)
\ee 
between BHDE and matter, the continuity equations~\eqref{c1} and~\eqref{c2} become
\begin{eqnarray}
\label{c3}
&&\dot\rho_m+3H\rho_m\,=\,Q\,,\\[2mm]
&&\dot\rho_D+3H\rho_D(1+\omega_D)\,=\,-Q\,.
\label{c4}
\end{eqnarray}
Following similar calculations as above, one can show that Barrow
holographic energy equation of state takes the form
\be
\omega_D\,=\,\frac{\Delta+2b^2/\Omega_D}{(2-\Delta)\Omega_D-2}\,,
\ee
where $b$ is the coupling parameter that quantifies the interaction. 

\subsection{The age of the Universe}
Let us now consider the following integral
\be
t=\int_0^{t_{Universe}} dt =\frac{2\left(\frac{C}{3 M_p^2}\right)^{1/\Delta}}{3\,\Delta}\int_{\Omega_{D,i}}^{\Omega_{D,0}} \frac{1-\left(1-\Delta/2\right)\Omega_D}{\Omega_D^{1-1/\Delta}\left(1-\Omega_D\right)}\,d\Omega_D\,,
\label{test}
\ee
{where $\Omega_{D,i}\equiv\Omega_D(t=0)$ and $\Omega_{D,0}$ is the current value of the DE density. }
Here, we have used Eqs.~\eqref{dotH2} and \eqref{omegaD} along with
\be
\frac{dH}{H^2}\,=\,-\frac{\left(\frac{C}{3 M_p^2}\right)^{1/\Delta}}
{\Delta\,\Omega_D^{1-1/\Delta}}
\,d\Omega_D\,.
\ee 

By integrating the above relation, it follows that
\be
t=3^{-\frac{2+\Delta}{\Delta}}\left(\frac{C^2}{M_p^4}\right)^{\frac{1}{\Delta}}\frac{1}{H(1+\Delta)}\left[2+2\Delta+\frac{C\Delta}{3M_p^2H^{\Delta}}\,{}_2 F_1(1,1+\frac{1}{\Delta};2+\frac{1}{\Delta};\frac{C}{3M_p^2\,H^{\Delta}})
\right]\Bigg|_{z=0}\,,
\ee
where ${}_2 F_1(a,b;c;d)$ is the hypergeometric function of first kind.
This equation can be used to estimate the order
of the age of the current Universe ($z=0$) in our model. 
Specifically, by using the relation between the Hubble and EoS parameters, we can approximately write 
\be
t\approx 
\frac{2-\Delta}{3H_0}\left(1-\frac{\omega_D(z=0)}{1+\omega_D(z=0)}\right).
\label{tfin}
\ee
For $\omega_D(z=0)=-2/3$, we then 
have $t=1/H_0$ for $\Delta=1$, corresponding
to the maximal deformation of the Bekenstein-Hawking area law.
As observed in~\cite{Tavayef:2018xwx}, 
further corrections to Eq.~\eqref{tfin} may arise due
to either different modifications of the horizon entropy or
other IR cutoffs.

\section{Tachyon scalar field as Barrow holographic dark energy in a non-flat FRW Universe}
\label{Tachnonflat}

In this Section we analyze the correspondence between
the tachyon dark energy model and the BHDE scenario in a non-flat FRW Universe. 
Toward this end, we recall that in~\cite{Setare:2007hq} it has been shown that
the energy density $\rho_T$ and pressure $p_T$ for the tachyon scalar field take the form
\begin{eqnarray}
\label{ted}
\rho_T&=&\frac{V(T)}{\sqrt{1-\dot T^2}}\,,\\[2mm]
p_T&=&-V(T)\sqrt{1-\dot T^2}\,,
\label{ped}
\end{eqnarray}
where $V(T)$ is the tachyon potential energy. From these relations,
we derive the equation of state parameter (EoS) for the tachyon as 
\be
\label{omegaT}
\omega_T\,=\,p_T/\rho_T=\dot T^2-1\,,
\ee 
which will be later equated to the EoS parameter of BHDE
in our reconstructed scenario. We also remind that 
the cosmological model based on the effective Lagrangian of tachyonic matter
\be
\label{action}
\mathcal{L}(T)=-V(T)\sqrt{1-T_{,\mu}T^{,\mu}}\,,
\ee
{with $V(T)=\emph{const.}$ coincides with Chaplygin gas model\footnote{{In the present naive picture, the question we ask ourselves is how BHDE would appear if one imposes that the evolution of its energy density can be described in terms of that of tachyon field. More rigorously, one should derive the scalar field equation by computing the variation of the action including the lagrangian~\eqref{action} with respect to the tachyon field T and specializing the result to the metric in Eq.~\eqref{FRW}. The ensuing expression of $\dot T$ could then be used to fix the dynamics of the model (and, in particular, the scale factor) by comparison with BHDE.}}}~\cite{Chap}. 

Let us explicitly explore the connection between BHDE and 
tachyon dark energy model. Toward this end, we consider the time derivative of BHDE~\eqref{brhod}  and use Eq.~\eqref{dotL} to get
\be
\dot\rho_D\,=\,C(\Delta-2)L^{\Delta-3}\left[1-\frac{1}{\sqrt{|k|}}\cos\mathrm{n}(\sqrt{|k|}x)\right],
\ee
where $x\equiv R_h/a$. By means of the continuity equation~\eqref{c4}, this can be cast as
\be
-3H\rho_D(1+\omega_D)-3b^2H(\rho_m+\rho_D)=C(\Delta-2)L^{\Delta-3}\left[1-\frac{1}{\sqrt{|k|}}\cos\mathrm{n}(\sqrt{|k|}x)\right].
\ee
We can now resort to Eq.~\eqref{omegaT}  to obtain
\be
-3 H \rho_D \dot T^2-3b^2H(\rho_m+\rho_D)=C(\Delta-2)L^{\Delta-3}\left[1-\frac{1}{\sqrt{|k|}}\cos\mathrm{n}(\sqrt{|k|}x)\right],
\ee
where we have made use of the required correspondence 
$\omega_T=\omega_D$, 
as explained above.
This relation can be further manipulated by dividing both sides
by $3H\rho_D$ and using Eq.~\eqref{u} to give
\be
-\dot T^2-b^2(u+1)=\frac{C(\Delta-2)L^{\Delta-3}}{3H\rho_D}
\left[1-\frac{1}{\sqrt{|k|}}\cos\mathrm{n}(\sqrt{|k|}x)\right].
\ee
After employing Eq.~\eqref{brhod} and the condition $L=H^{-1}$, 
we finally reach
\be
-\dot T^2-b^2(u+1)=\frac{\Delta-2}{3}
\left[1-\frac{1}{\sqrt{|k|}}\cos\mathrm{n}(\sqrt{|k|}x)\right],
\ee
which can be equivalently written as
\be
\dot T^2=\frac{2}{3}-b^2(u+1)-\frac{\Delta}{3}+\,\frac{\Delta-2}{3}\frac{1}{\sqrt{k}}\cos\mathrm{n}(\sqrt{|k|}x)\,.
\ee
Notice that, once known $\dot T^2$, one can express
the tachyon potential $V(T)$ in terms of $H$ and the parameter $\Delta$
by simply equating Eqs.~\eqref{brhod} and~\eqref{ted}. 
 
\begin{figure}[t]
\begin{center}
\includegraphics[width=10.5 cm]{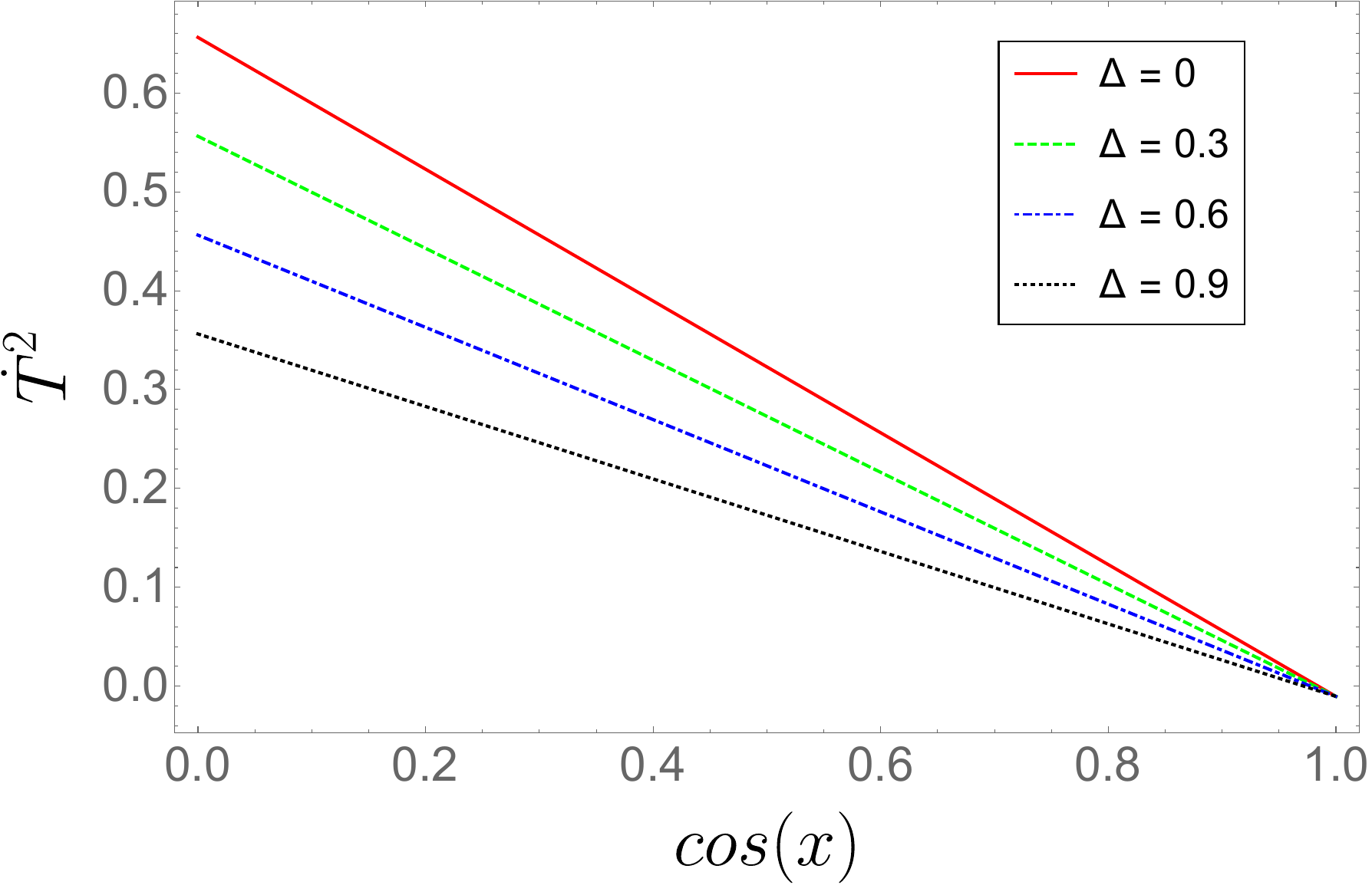}
\caption{Evolution trajectories of $\dot T^2$ for a closed ($k=1$) Universe. We set $u=0.04$ and $b^2=0.01$ as in~\cite{Liu:2021heo}.}
\label{fig1}
\end{center}
\end{figure}

Let us now study the cases $k=0, 1,-1$  separately. 
The $k=0$ framework corresponds to a flat FRW Universe.
In this case it is easy to show that 
\be
\label{tbdot}
\dot T^2\,=\,-\frac{b^2}{\Omega_D}\,,
\ee
where we have used Eqs.~\eqref{cosn} and~\eqref{u}. 
Thus, if we require $\dot T$ to be real, then $b=0$ in flat space, 
which allows us to conclude that $\dot T^2$ must always be vanishing in flat space, independently of the existence of interaction.
In turn, from Eqs.~\eqref{ted} and~\eqref{ped} this implies 
\be
\rho_T=-p_T\,.
\ee
We conclude that in this case, the equation of state is always $\omega_D=-1$, reproducing a cosmological constant-like behavior.

On the other hand, for $k=1$ (closed FRW Universe) we get from the definition~\eqref{cosn}
\be
\label{k1}
\dot T^2\,=\,\frac{2}{3}-b^2(u+1)-\frac{\Delta}{3}+\frac{\Delta-2}{3}\cos(x)\,.
\ee
In order for $\dot T$ to be zero (i.e. $\omega_T=-1$), we must have
\be
\label{cosxT0}
\cos(x)\,=\,\frac{3b^2(u+1)}{\Delta-2}+1\,,
\ee
which clearly admits non-trivial solution, since $\Delta<1$. 
This provides the condition for which the tachyon model of BHDE reproduces the cosmological constant behavior
in a closed FRW Universe. By contrast, in~\cite{Liu:2021heo} it is
argued that $\dot T^2$ cannot be zero in Tsallis holographic dark energy in a non-flat Universe. 

The evolution trajectory of $\dot T^2$ in Eq.~\eqref{k1} is plotted in Fig.~\ref{fig1} for fixed $b$ and $u$ and various allowed $\Delta$. One can see that $\dot T^2$ decreases monotonically for increasing $\cos(x)$ and $\dot T^2\ge0$, so that $\omega_T\ge-1$ (quintessence or cosmological constant-like behavior), while tending to negative values for $cos(x)\rightarrow1$, which becomes physically invalid.
The non-interacting case can be simply derived
by setting $b^2=0$ in Eqs.~\eqref{k1} and~\eqref{cosxT0}.
The ensuing behavior is similar to that described above, 
the only difference being that $\dot T^2\ge0$ throughout the whole
evolution in this case.

Similarly, we can consider the dynamics of the
tachyon field in an open ($k=-1$) Universe. Following 
the same reasoning as above, we obtain
\be
\label{kmeno1}
\dot T^2\,=\,\frac{2}{3}-b^2(u+1)-\frac{\Delta}{3}+\frac{\Delta-2}{3}\cosh(x)\,.
\ee

\begin{figure}[t]
\begin{center}
\includegraphics[width=10.5 cm]{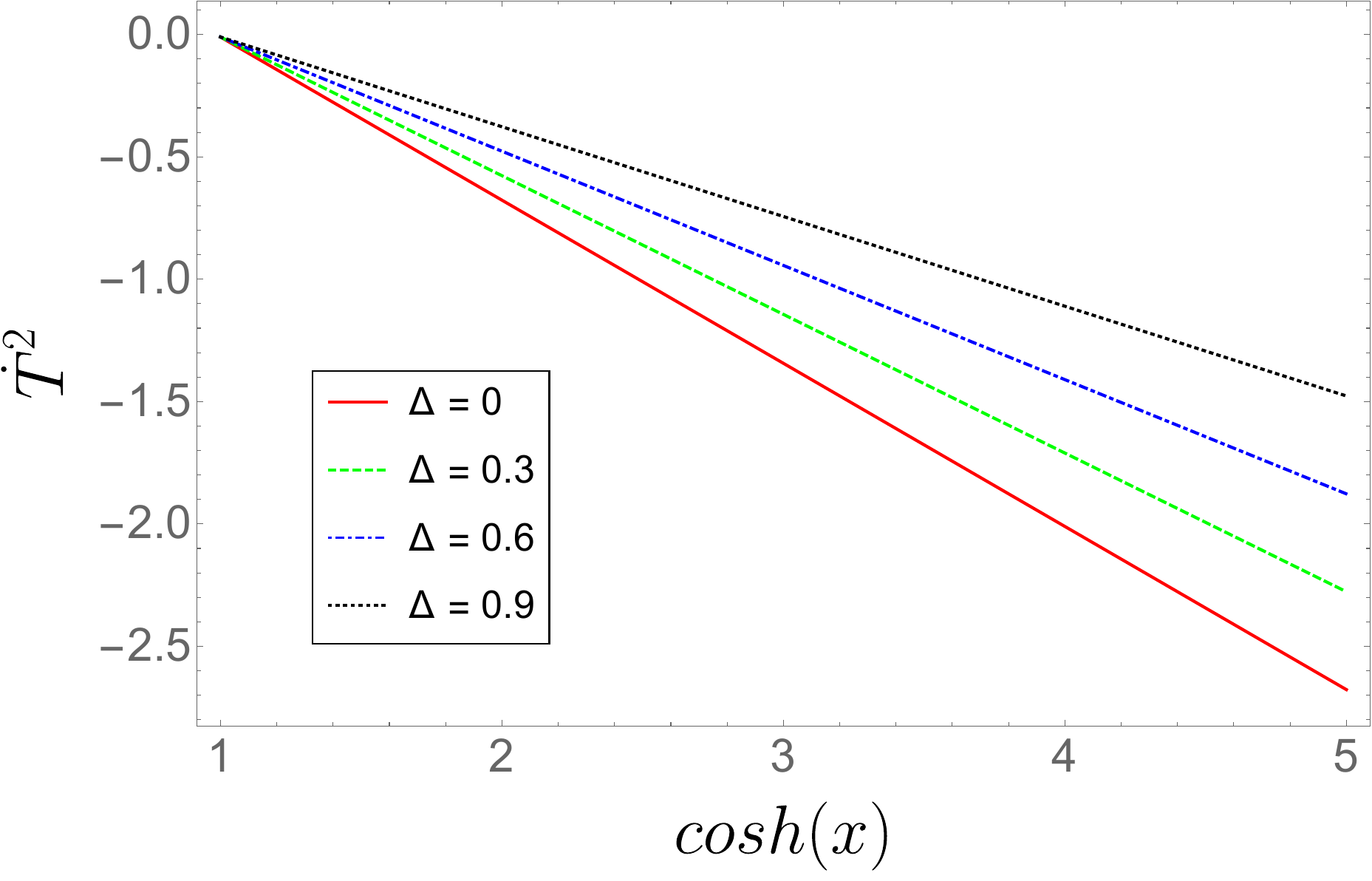}
\caption{Evolution trajectories of $\dot T^2$ for an open ($k=-1$) Universe. We set $u=0.04$ and $b^2=0.01$ as in~\cite{Liu:2021heo}.}
\label{fig2}
\end{center}
\end{figure}

The evolution of $\dot T^2$ in Eq.~\eqref{kmeno1} is
plotted in Fig.~\ref{fig2}  for different values of $\Delta$.
As before, we notice that $\dot T^2$
decreases monotonically for increasing $\cosh(x)$, 
but in this case it is always negative, 
which is not a physically valid situation. 
We also see that Eq.~\eqref{kmeno1} vanishes, provided that
\be
\label{cosh}
\cosh(x)\,=\,\frac{3b^2(u+1)}{\Delta-2}+1\,.
\ee
However, since $\cosh(x)>1$ (we remind that $x\neq0$), we infer $\dot T^2$ can never be zero, in agreement with the result of~\cite{Liu:2021heo}. The same behavior is exhibited in the absence of interactions ($b^2=0$).

\subsection{Observational studies}
\begin{figure}[t]
\begin{center}
\includegraphics[width=10.5cm]{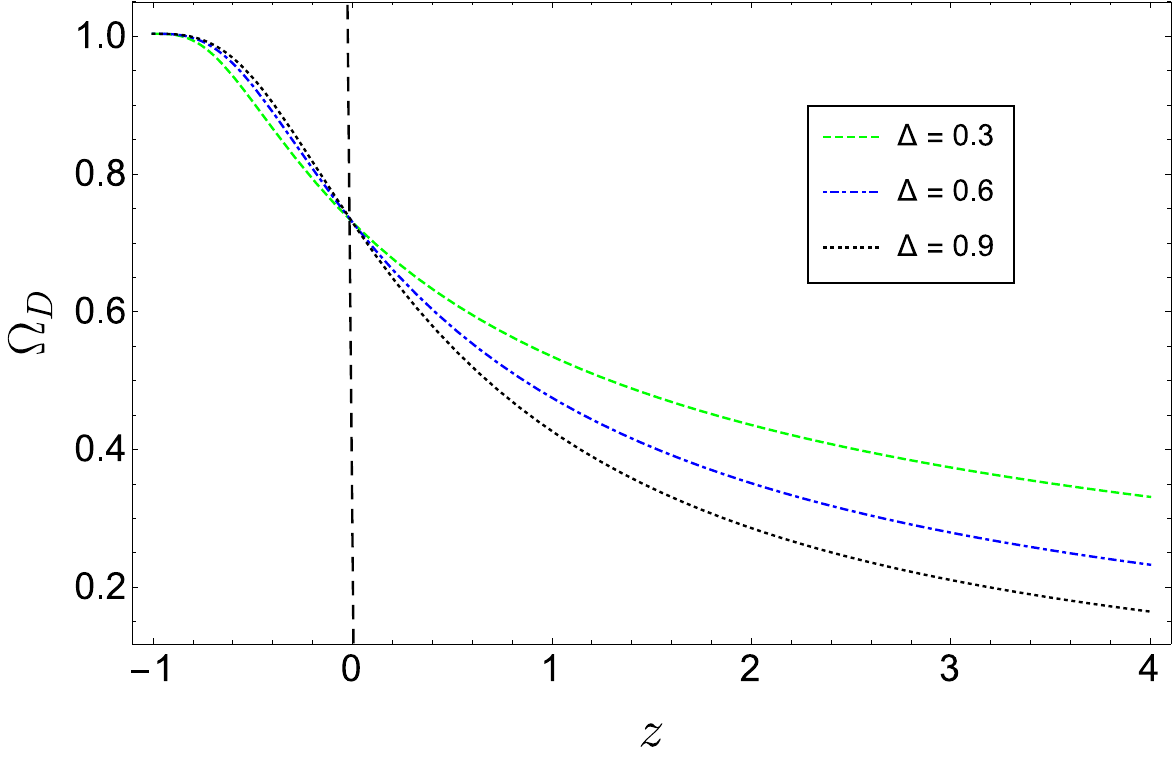}
\caption{Evolution trajectories of $\Omega_D$ versus $z$ (we set $\Omega_k=0.01$ and $\Omega_D^0=0.73$). The dashed vertical line marks the value at present time.}
\label{Fig3}
\end{center}
\end{figure}

\begin{figure}[t]
\begin{center}
\includegraphics[width=10.5 cm]{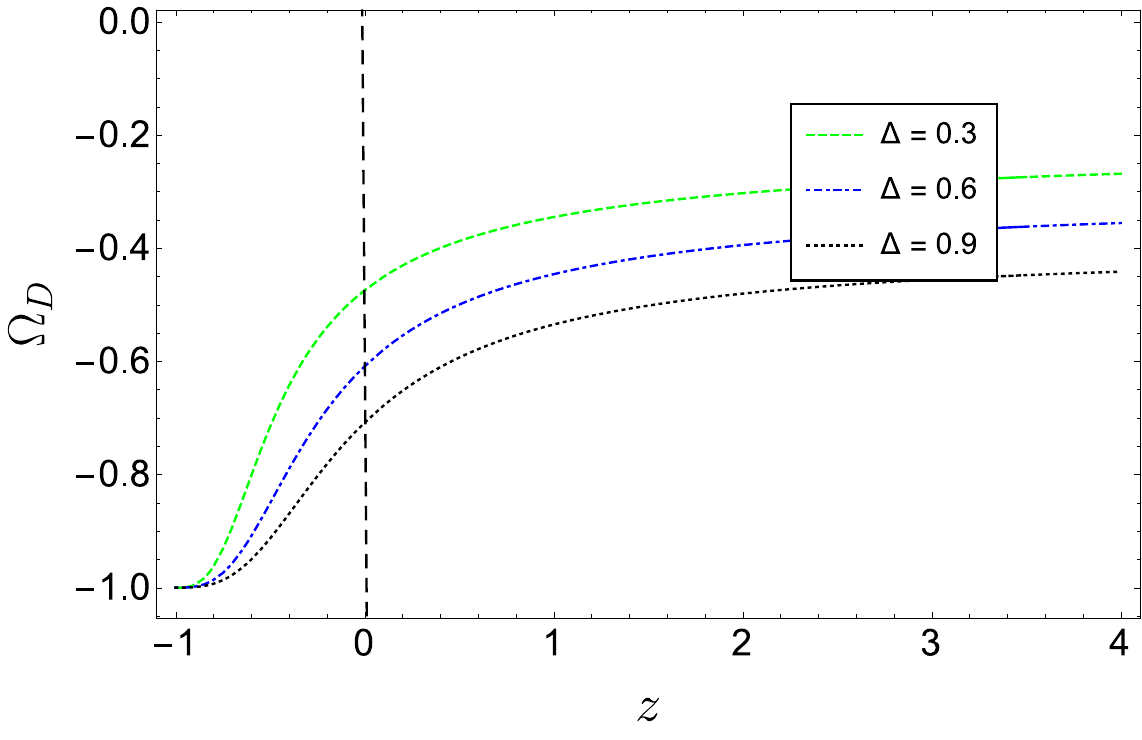}
\caption{Evolution trajectories of $\omega_D$ versus $z$ (we considered the same initial conditions as in Fig.~\ref{Fig3}). The dashed vertical line marks the value at present time.}
\label{Fig4}
\end{center}
\end{figure}

{This section is devoted to explore some observational implications
of BHDE. For simplicity, we focus on the case where there is no interaction between the dark sectors of the cosmos and positive values of curvature. 
First, we notice that the use of Eq.~\eqref{dotH1} in non-flat Universe
allows us to write down the deceleration parameter $q$ as
\be
q=-\frac{\ddot a}{aH^2}=-1-\frac{\dot H}{H^2}=-1-\Omega_k+\frac{3\Omega_D}{2}\left(1+u+\omega_D\right).
\label{qpar}
\ee
From the above definition, one can check that
$q>0$ ($q<0$) corresponds to a decelerated (accelerated) expansion 
of the Universe, since $\ddot a<0$ ($\ddot a>0$).
By computing the time derivative of Eq.~\eqref{brhod}, we then
get
\be
\label{dotrho}
\dot\rho_D=\left(2-\Delta\right)\rho_D\,H\left[\Omega_k-\frac{3\Omega_D}{2}\left(1+u+\omega_D\right)\right], 
\ee
while the further usage of the definition~\eqref{Om} leads to
\be
\label{53}
\Omega_D'=-\Omega_D\left[\frac{1}{2}\left(2-\Delta\right)\left(3+\Omega_k+3\Omega_D\,\omega_D\right)\right.\,+\,2\Omega_k-3\Omega_D\left(1+u+\omega_D\right)
\bigg].
\ee
where $\Omega'_D\equiv\frac{d\Omega_D}{d(\log a)}$.
By plugging into~\eqref{c2}, the BHDE EoS parameter and the fractional BHDE density take the form
\begin{eqnarray}
\label{EoSparamom}
\omega_D&=&-\frac{3+\left(\frac{\Delta}{2}-1\right)\left(\Omega_k+3\right)}
{3\left[1+\left(\frac{\Delta}{2}-1\right)\Omega_D\right]}\,,\\[3mm]
\Omega_D'&=&\frac{\Omega_D\,\Delta}{2}\,\frac{3+\Omega_k-3\Omega_D}{1+\left(\frac{\Delta}{2}-1\right)\Omega_D}\,.
\end{eqnarray}
The latter equation is solved numerically and plotted in Fig.~\ref{Fig3}, which shows a monotonic increasing of dark energy
for decreasing redshift, implying a DE dominated Universe in the far future. On the other hand, the evolution of the EoS parameter is displayed in Fig.~\ref{Fig4}, which indicates that BHDE behaves like quintessence or (asymptotically) cosmological constant, consistently with the discussion below Fig.~\ref{fig2}. Specifically, for the considered $\Delta$'s, we can see that
the present value of this parameter lies in the range 
$-0.7\lesssim\omega_{D,0}\lesssim-0.4$, which only slightly deviates
from recent Planck+WP+BAO predictions~\cite{Planck}.}

\begin{figure}[t]
\begin{center}
\includegraphics[width=10.5cm]{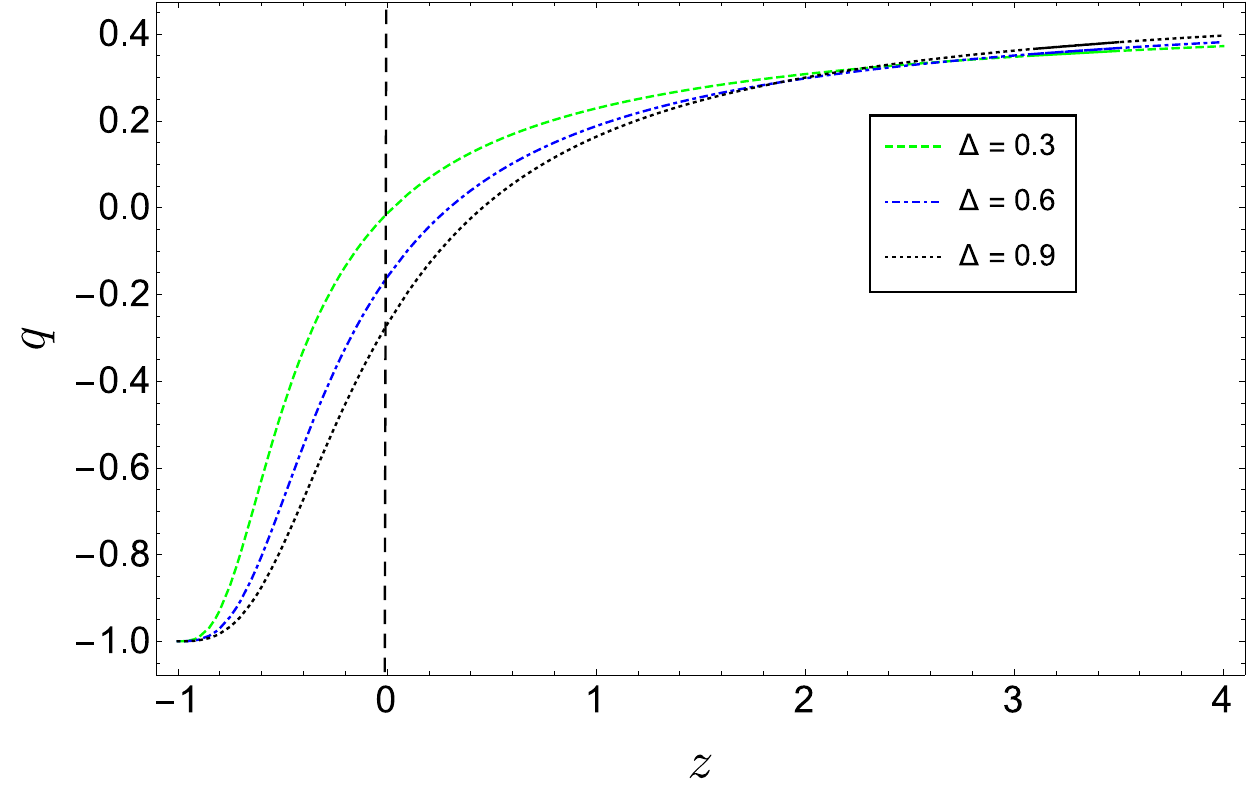}
\caption{Evolution trajectories of $q$ versus $z$ (we considered the same initial conditions as in Fig.~\ref{Fig3}). The dashed vertical line marks the value at present time.}
\label{Fig5}
\end{center}
\end{figure}

{Similarly, from Eq.~\eqref{qpar} we can derive the evolution of the deceleration parameter, whose behavior is plotted in Fig.~\ref{Fig5}.
We observe that  the present model  allows to explain the 
sequence of an early decelerating (i.e. $q>0$) expansion of the Universe, followed by an accelerated (i.e. $q<0$)
epoch. Also, it is consistent with the description of the current
acceleration, although the predicted value $-0.3\lesssim q_0\lesssim-0.01$ is slightly higher than $q_0\simeq-0.5$
obtained in the standard $\Lambda$CDM model~\cite{Planck}}. 

{To further check the phenomenological consistency of our model, one could also study the evolution of 
some other relevant quantities, such as
the Hubble rate or the effects of Barrow model on the growth of cosmological perturbations and structure formation. The analysis of 
these aspects is more challenging and requires further effort. Results will be presented in a future work. }

\section{Inflation in Barrow Holographic Dark Energy}
\label{infl}
In this Section we discuss inflation in BHDE. For reasons that will appear clear below and following~\cite{Inft}, here we consider the more general expression for the length scale $L^{-2}=\alpha H^{2}+\beta\dot{H}$, where $\alpha$ and $\beta$ are dimensionless constant. 
Assuming that the expansion of the Universe
is driven by BHDE~\eqref{brhod} and neglecting the matter
contribution due to the rapid inflationary expansion, Eq.~\eqref{ffe}
becomes
\be
\label{neglect}
H^2\,=\,\frac{C}{3M_p^2}\left(\alpha H^2+\beta\dot H\right)^{1-\Delta/2}\,,
\ee
from which we infer
\be
\label{tderiv}
\dot H\,=\,\frac{H^2}{\beta}\left[\left(\frac{3M_p^2}{C}\right)^{\frac{2}{2-\Delta}}(H^2)^{\frac{\Delta}{2-\Delta}}-\alpha
\right].
\ee
From this relation, it is clear that setting 
the IR cutoff $L\simeq H^{-1}$ (i.e. $\beta=0$) as in the previous study 
would give rise to technical issues in the present framework.

To simplify the resolution of Eq.~\eqref{tderiv}, we introduce
the e-folds variable $N=\log\left(a/a_i\right)$, where $a_i$ is the
initial value of the scale factor $a$. 
By observing that $dN=Hdt$ and $\dot H=\frac{1}{2}\frac{dH^2}{dN}$, 
integration of Eq.~\eqref{tderiv} gives
\be
\label{logH}
\log\left\{\tilde H^2\left[\gamma\left(\tilde H^2
\right)^{\frac{\Delta}{2-\Delta}}
\right]^{1-2/\Delta}
-\alpha
\right\}\Bigg|_{\tilde H_i}^{\tilde H_f}\,=\,-\frac{2\alpha N}{\beta}\,,
\ee
where $\tilde H=H/M_p$ is the dimensionless Hubble parameter and
\be
\label{gamma}
\gamma=\left(\frac{3M_p^2}{C}\right)^{\frac{2}{2-\Delta}}\,M_p^{\frac{2\Delta}{2-\Delta}}\,. 
\ee 
Here we have denoted the Hubble parameter
at the end of inflation by $\tilde H_f$. 

From Eq.~\eqref{neglect} we can now compute
the characteristic parameters of slow-roll inflation. Specifically, the 
first slow-roll parameter is given by
\be
\label{ep1}
\epsilon_1\,=\,-\frac{\dot H}{H^2}=-\frac{1}{\beta}\left[\gamma\left(\tilde H^2
\right)^{\frac{\Delta}{2-\Delta}}-\alpha\right].
\ee
The other  slow-roll parameters can be derived
by using the definition $\epsilon_{n+1}=d\log(\epsilon_n)/dN$.
For the second parameter $\epsilon_2$ we get
\be
\label{ep2}
\epsilon_2\,=\,\frac{\dot \epsilon_1}{H\epsilon_1}=\frac{2\gamma}{\beta}\left(\frac{\Delta}{2-\Delta}\right)\left(\tilde H^2
\right)^{\frac{\Delta}{2-\Delta}}\,.
\ee

Let us now evaluate the Hubble parameter at the end of inflation.
This phase is characterized by $\epsilon_1=1$.
By straightforward calculations, we obtain
\be
\label{Hf}
\tilde H_f^2\,=\,\left(\frac{\gamma}{\alpha-\beta}\right)^{1-2/\Delta}\,.
\ee
On the other hand, at the beginning of inflation (including
the horizon crossing time) Eq.~\eqref{logH} gives
\be
\tilde H_i^2\,=\,\left[\frac{\gamma}{\alpha}\left(1+\frac{\beta\,e^{2\alpha N/\beta}}{\alpha-\beta}\right)\right]^{1-2/\Delta}\,,
\ee
which can be used to calculate the slow-roll parameters for earlier time
by direct substitution in Eqs.~\eqref{ep1} and~\eqref{ep2}.

In order to derive the scalar spectral index $n_s-1$
and the tensor-to-scalar ratio $r$, we follow~\cite{PLBSar,Inft}
and make use of the usual perturbation procedure. We are led to
\be
\label{nsr}
n_s-1 \,=\,-2\epsilon_1-2\epsilon_2\,,\qquad r=16\epsilon_1\,.
\ee
Clearly, a full perturbation analysis is needed to 
obtain the exact expressions of $n_s-1$ and $r$. 

Two comments are in order here: first, we notice
that the constant $\gamma$ does not intervene
in the calculation of the slow-roll parameters at 
the horizon crossing time, which means that neither
$n_s-1$ nor $r$ depend on it. As explained in~\cite{Inft}, 
this constant can be estimated by considering the amplitude
of the scalar perturbation.
Furthermore, it is worth mentioning that a similar analysis of inflation and correspondence between BHDE and tachyon field has been proposed in~\cite{Prelimin}. However, in that case the authors consider 
values of Barrow parameter 
$\Delta$ higher than unity, which are actually forbidden
in Barrow model. This somehow questions the results exhibited in~\cite{Prelimin}.

\subsection{Trans-Planckian Censorship Conjecture}
The large-scale structures we currently see in the Universe
originated from matter and energy quantum fluctuations 
produced during inflation. Such fluctuations 
cross the Hubble radius during the early phase,
are stretched out and classicalize, and finally 
re-enter the Hubble horizon to produce the CMB anisotropies.
The key point is that if inflation lasted longer than  
the supposed minimal period, 
then it would be possible to observe length scales 
originated from modes smaller than the Planck length
at inflation~\cite{Martin}. This problem is usually referred to
as ``Trans-Planckian problem''. To avoid inconsistencies, 
it has been conjectured that this problem cannot arise
in any consistent model of quantum gravity (``Trans-Planckian Censorship Conjecture'', TCC)~\cite{Bedroya}.

The TCC states that no length scales which cross the
Hubble horizon could ever have had a wavelength smaller
than the Planck length. This is  imposed 
by requiring that 
\be
\label{TCC}
\frac{L_p}{a_i} \,<\,\frac{H_f^{-1}}{a_f}\,,
\ee
where $L_p=1/M_p$ is the Planck length and we have
denoted by $a_f$  the scale factor at the end of inflation.
By using Eq.~\eqref{Hf} for the Hubble parameter
at the final time, the TCC~\eqref{TCC} becomes
\be
\left(\frac{\gamma}{\alpha-\beta}\right)^{1-2/\Delta}<(8\pi\hspace{0.2mm}e^{N})^2\,,
\ee
the validity of which can be examined by comparison with 
observational data. This aspect is under active investigation
and will be addressed in more detail in a future work.

\section{Conclusions and outlook}
\label{Conc}

The origin of the accelerated expansion
of the Universe is an open problem
in modern Cosmology. To date, the most reliable explanation
is provided by the existence of an enigmatic form of energy - the Dark Energy - affecting the Universe on large scales. Several candidates
have been considered to account for this phenomenon. 
In particular, the holographic dark energy has been 
largely studied, also in connection with different
real scalar field theories, such as quintessence~\cite{Ratra:1987rm,Frieman:1995pm,Turner:1997npq,Caldwell:1997ii}, K-essence~\cite{Armendariz-Picon:2000nqq},
phantom~\cite{Caldwell:1999ew,Caldwell:2003vq,Nojiri:2003vn}, interacting models~\cite{Deffayet:2001pu} and tachyon~\cite{Setare:2007hq} (see also~\cite{colg1,colg2} for arguments that theoretically rule out quintessence and K-essence). Recently, the interest has been extended to
the Tsallis holographic dark energy~\cite{Tavayef:2018xwx,Saridakis:2018unr,Saridakis:2020zol,Moradpour:2020dfm,Drepanou:2021jiv} and the possibility of using it to describe the dynamics of
the tachyon field~\cite{Liu:2021heo}. 

In this work we have considered the further scenario
of tachyon model as Barrow holographic dark energy.
Barrow entropy  arises from the effort to include
quantum gravity effects on the black hole horizon. In 
this sense, the present analysis must be intended
as a preliminary step toward a fully quantum gravity extension of~\cite{Liu:2021heo}. In the absence of empirical guidelines, 
we have exploited the powerful tool of Lagrangian formalism
and, in perspective, the precious clues that may 
be provided by the related Noether's theorem.
In particular, we have established a correspondence between
BHDE and the tachyon field model in a FRW Universe, 
both in the presence and absence of interactions between dark energy and matter. {In this regard, we would like to stress that the aim of the present analysis is to reconstruct BHDE in tachyon model, rather than showing the consistency between the two frameworks. In other terms, in the absence of a lagrangian formulation of BHDE, the question we have asked ourselves is how BHDE would appear if one requires that the evolution of its energy density can be described in terms of that of tachyon field. As a result, we have shown that the tachyon field should exhibit a non-trivial dynamics. 
In particular, in a flat Universe, $\dot T^2$ must always be vanishing, 
independently of the existence of interaction, which implies $\omega_D=-1$ for the equation-of-state parameter. On the other hand,  for a non-flat Universe and various values of Barrow parameter, we have found that $\dot T^2$ decreases monotonically for increasing $\cos(R_h/a)$ and $\cosh(R_h/a)$. Specifically, $\dot T^2\ge0$ for a closed Universe, while $\dot T^2<0$ for an open one, which is physically not allowed. We have finally investigated an inflationary scenario described by a Universe filled with BHDE and commented on the Trans-Planckian Censorship Conjecture.}

We have shown that the tachyon field exhibits a non-trivial dynamics. 
In a flat Universe, $\dot T^2$ must always be vanishing, 
independently of the existence of interaction, which implies $\omega_D=-1$ for the equation-of-state parameter. On the other hand,  for a non-flat Universe and various values of Barrow parameter, we have found that $\dot T^2$ decreases monotonically for increasing $\cos(R_h/a)$ and $\cosh(R_h/a)$. Specifically, $\dot T^2\ge0$ for a closed Universe, while $\dot T^2<0$ for an open one, which is physically not allowed. We have finally discussed observational consistency of our model and investigated an inflationary scenario described by a Universe filled with BHDE, and commented on the Trans-Planckian Censorship Conjecture.

Further aspects remain to be addressed. For instance, 
we can look at the correspondence between the tachyon field and other dark energy scenarios, in particular stable dark energy models.
Furthermore, it would be interesting to disclose the effects of Barrow entropic corrections on the growth of perturbations and structure formation in the present model. Preliminary studies along this direction have been recently conducted in~\cite{Sheykhi:2022gzb}.
{One more suggestive perspective concerns the extension of the above framework by using different IR cutoffs} (such as the future event horizon or Grand-Oliveros cutoff) and/or other deformed entropies, such as Kaniadakis entropy~\cite{Drepanou:2021jiv}, which is based on a relativistic self-consistent generalization of the classical Boltzmann-Gibbs entropy~\cite{Kaniadakis:2002zz} (see also~\cite{RevLuc} for a recent review of Kaniadakis entropy applications in gravity and cosmology). 
Finally, it is essential to examine to what extent
our effective model reconciles with predictions of 
more fundamental candidate theories of quantum gravity,  
such as String Theory and Loop Quantum Gravity, or more phenomenological approaches, such as generalizations of the Heisenberg relation at Planck scale~\cite{Kempf:1994su,Scardigli:1999jh,Bosso:2018syo,Luciano:2021cna}.
Work along these directions requires further investigation
and will be presented elsewhere.

\acknowledgments
G.~G.~L. acknowledges the Spanish ``Ministerio de Universidades'' 
for the awarded Maria Zambrano fellowship and funding received
from the European Union - NextGenerationEU. He also acknowledges 
participation in the COST Association Action CA18108  ``Quantum Gravity Phenomenology in the Multimessenger Approach'' and LISA Cosmology Working group. He is finally grateful to Pasquale Bosso for technical suggestions on the final editing of the manuscript.

\bibliography{Bib}


%
%

\end{document}